\documentclass[prb,twocolumn,showpacs,preprintnumbers,amsmath,amssymb,floatfix,able]{revtex4}

\usepackage{graphicx, epsfig}
\usepackage{xcolor}
\usepackage{bm}
\usepackage{amssymb}
\usepackage{amsmath}
\usepackage{amsfonts}
\usepackage{hyperref}
\usepackage{array}

\newcommand{\fgf}{FC$_2$F}
\renewcommand{\gg}{C$_8$}
\newcommand{\gf}{F$_4$/C$_8$}
\newcommand{\hh}{H$_2$}
\newcommand{\oo}{O$_2$}
\newcommand{\of}{(OF)$_2$}
\newcommand{\oh}{(OH)$_2$}
\newcommand{\nn}{N$_2$}
\newcommand{\ff}{F$_2$}
\newcommand{\f}{F}
\newcommand{\fff}{F$_3$}
\newcommand{\xe}{Xe}
\renewcommand{\deg}{$\rm^{\,o}C$}
\newcommand{\super}{\rm F_{16}/C_{32}/F_{14}}

\begin{document}

\title{Novel First-Principles Insights into Graphene Fluorination}

\author{Tahereh Malakoutikhah}
\email{t.malakoutikhah@gmail.com}
\author{S. Javad Hashemifar} 
\email{hashemifar@iut.ac.ir }
\author{Mojtaba Alaei} 
\affiliation{ Department of Physics, Isfahan University of
Technology, Isfahan 84156-83111, Iran}


\begin{abstract}

Comprehensive first-principles calculations are performed on diverse
arrangements of relevant chemical defects in fluorographene to provide accurate
microscopic insights into the process of graphene fluorination. 
The minimum energy paths for the half- and full-fluorination processes
are calculated for a better understanding of these phenomena. 
While experimental observations indicate a much slower rate of the full-fluorination process,
compared with the half-fluorination one, the obtained energy profiles demonstrate
much enhanced fluorine adsorption after the half-fluorination stage.
This ambiguity is explained in terms of significant chemical activation 
of the graphene sheet after half-fluorination,
which remarkably facilitates the formation of chemical contaminants in the system
and thus substantially slows down the full-fluorination procedure.
After considering the binding energy and durability of the relevant chemical species,
including hydrogen, oxygen, and nitrogen molecules and xenon atom,
it is argued that oxygen-fluorine ligands are the most likely chemical contaminants 
opposing the full-fluorination of a graphene sheet.
We propose an oxygen desorption mechanism for the atomic description 
of the full-fluorination procedure in realistic situations.
It is argued that the proposed mechanism explains well much enhanced
rate of the full-fluorination procedure at elevated temperatures.
\end{abstract}

\maketitle

\section{Introduction}

Fluorographene (FG) is a distinguished member of graphene (Gr) derivatives, 
which is formed by covering both sides of a Gr sheet with a fluorine monolayer, 
leading to the FC$_2$F chemical formula. 
The applicable properties of FG added to its extraordinary 
features inherited from pristine graphene, 
have attracted great interest over the last decade. 
The stoichiometric FG has an ordered structure that is thermally 
stable up to 400\deg.\cite{nair2010}
The experimental Young modulus of this 2D material is 100\,N/m 
which shows its high mechanical stability.
Finally, what makes FG prominent among others 
is having the widest measured band gap of about 3\,eV
among all of the Gr derivatives.\cite{nair2010}
Recent studies confirm its potential applications in batteries and supercapacitors,
electrochemical sensors, solar cells, organic field-effect transistors,
electrocatalytic oxygen and hydrogen reactions, spintronic devices,
anti-corrosion and self-cleaning coatings,
oil-water separation, biomedicine, optoelectronic, and photonic.
\cite{chronopoulos2017}
 
Fluorination of graphene sheets with a proper chemical agent 
is a systematic method for the synthesis of stoichiometric FG samples.
Xenon difluoride (XeF$_2$) is a popular fluorinating agent that is utilized 
for the fluorination of graphene at various temperatures.
Nair and colleagues observed that complete fluorination may 
take more than two weeks at 70\deg, while at 200\deg\ about 10\,hours
is enough to obtain full-fluorinated samples.\cite{nair2010}
Creutzburg et al. performed the same procedure at 120\deg\
and found that after about 48\,hours the Gr samples are 
fully fluorinated.\cite{creutzburg2021}
Kashtiban and others employed accurate electron microscopies to distinguish 
two stages in the fluorination procedure.
They argue that in the first stage, one side of the Gr sheet is 
rapidly fluorinated (half-fluorination stage)
and then the fluorination of the opposite side will be started
(full-fluorination stage).\cite{kashtiban2014}
These experimental results suggest the presence of some nontrivial 
and temperature-dependent mechanism against the adsorption of 
fluorine atoms on the graphene sheet, in the second fluorination stage. 
A theoretical molecular dynamics (MD) simulation further complicates the situation, 
as it shows that adsorption of a fluorine atom on a Gr sheet encourages adsorption 
of a second fluorine atom at the opposite side of the sheet.\cite{paupitz2012}
Another discrepancy is the theoretical band gap of FG 
which is much higher than the measured one. 
While the experimental band gap of FG is expected to be 
in the range of 3-3.8\,eV,\cite{nair2010,jeon2011}
the most accurate first-principles techniques predict 
a band gap of 5.1-7.49\,eV for this 2D system.
\cite{sahin2011,leenaerts2010,wei2013,yuan2015}
Considering some kinds of defects may somewhat reduce 
the theoretical band gap of FG toward the experimental value.
\cite{yuan2015,huang2017}

The nontrivial ambiguities observed in fluorographene
suggest possible misunderstanding in the graphene fluorination process.
Hence, in this work, we employ accurate first-principles calculations
to provide new microscopic insights for a more precise understanding
of the graphene fluorination procedure.

\section{Method}

Our electronic structure calculations and structural relaxations were performed 
in the framework of density functional theory by using the full potential 
numeric atom-center orbital (NAO) technique implemented in the FHI-aims package.\cite{fhiaims}
The NAO basis functions:
\begin{equation}
\varphi_i(r)=\frac{u_i(r)}{r} Y_{lm}(\omega)
\label{phi}
\end{equation}
are composed of radial solutions $u_i(r)$ of a Schr\"odinger-like equation 
and spherical harmonics $Y_{lm}$. 
The numerically tabulated functions, $u_i(r)$, are proven to be very 
efficient for accurate simulation of non-periodic and periodic systems.

For simulating the 2D sheets, we use a slab supercell containing 
a vacuum thickness of about 15\,\AA,
to prevent artificial interactions between adjacent layers.
Among possible atomic configurations for FG, we adopt the chair structure,
sketched in Fig.~\ref{struct}, which is argued to be the lowest energy
configuration of this 2D material.
A mesh of $8\times8\times1$ k-points was used for our Brillouin zone integrations
and the PBEsol exchange-correlation functional was employed 
for our structural optimizations.\cite{perdew2008} 
The scalar relativistic atomic-ZORA approximation (zero-order regular approximation) 
was implemented to involve the relativistic effects,\cite{zora}
while the full relativistic spin-orbit effect was only 
considered for the adsorption of the heavy Xe element.   
Moreover, the van der Waals correction based on the Hirshfeld partitioning of 
the electron density and spin polarization was considered 
in the calculations.\cite{tkatchenko2012}

\section{Results and Discussions}
 
\begin{figure}
\includegraphics[scale=0.105]{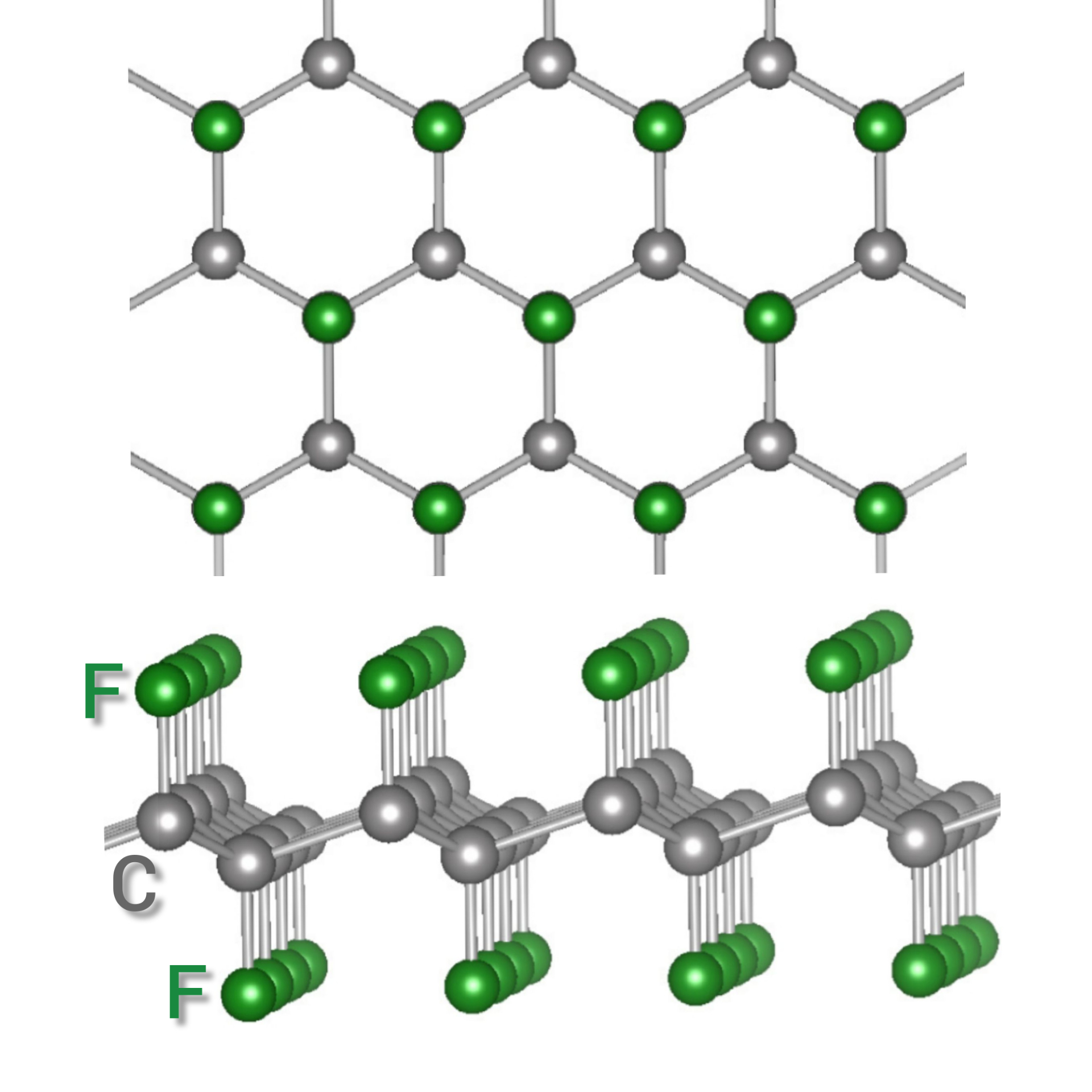}
\includegraphics[scale=0.95]{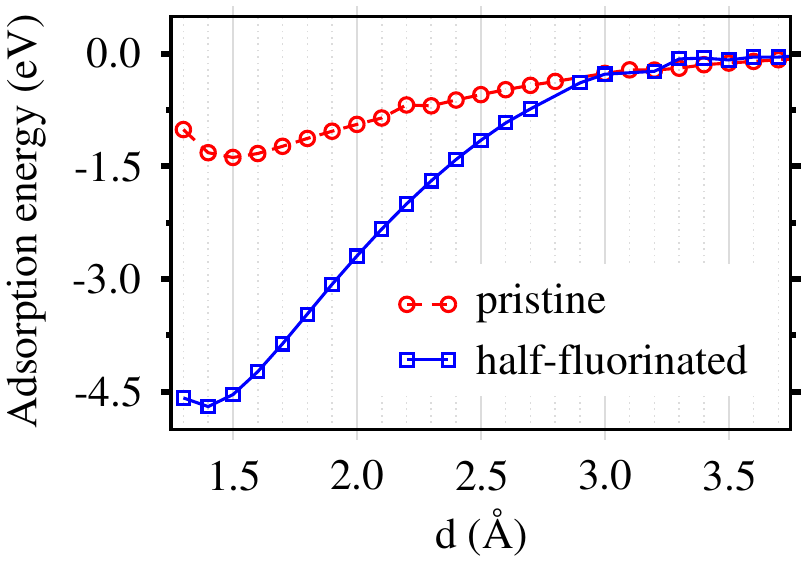}
\caption{\label{path}
 Top and side views of the chair structure of a fluorographene monolayer (\fgf)
 and the calculated adsorption energy path of a fluorine atom
 on a pristine graphene sheet and a half-fluorinated one. 
}
\end{figure}

As it was mentioned in the Introduction, a major experimental observation in the FG synthesis 
is that fluorination of the second side of the system takes weeks to be completed \cite{kashtiban2014} 
and it is significantly accelerated by increasing temperature 
from 70 to 200\deg.\cite{nair2010}
This observation may be attributed to the appearance of an energy barrier for the adsorption 
of fluorine atoms on the opposite side of a half-fluorinated graphene (hFG) sheet. 
Hence, in the first step, we calculated and compared the adsorption energy path of 
a fluorine atom on a pristine and a half-fluorinated Gr sheet.
In this regard, the energy of the system was minimized at several F atom-sheet distances
by constrained relaxation of the atomic positions and unit cell parameters,
in a $2\times2$ slab supercell.
The obtained energy path diagrams, presented in Fig.~\ref{path}, 
indicate no energy barrier for the adsorption of the fluorine atom on either 
the pristine or the half-fluorinated Gr sheet.
Moreover, the adsorption energy of the fluorine atom on hFG 
is remarkably enhanced with respect to the pristine sheet, 
in agreement with a recent MD simulation.\cite{paupitz2012}
These observations sound inconsistent with the very low experimentally 
observed rate of the full-fluorination process of graphene.\cite{nair2010}
Therefore, we propose a different scenario to explain this discrepancy.

The much higher binding energy of a fluorine atom on hFG, 
compared with the pristine Gr, indicates that the adsorption of 
a fluorine layer on a graphene sheet significantly activates 
the opposite facet of the sheet for adsorbing chemical species. 
Therefore, the opposite facet of the hFG sheet is expected 
to be suddenly contaminated with available chemical species in the environment. 
In other words, on the opposite side of hFG, close competition 
is expected between the fluorine atoms and other available chemical species 
to occupy the free adsorption sites of the sheet. 
This competition is anticipated to be the main mechanism for slowing 
the rate of the full-fluorination of a Gr sheet. 
Moreover, the presence and persistence of unwanted chemical species in the adsorption sites 
of FG may explain the large difference between the theoretical and 
experimental band gap of the system.

\begin{figure}
 \includegraphics[scale=0.1]{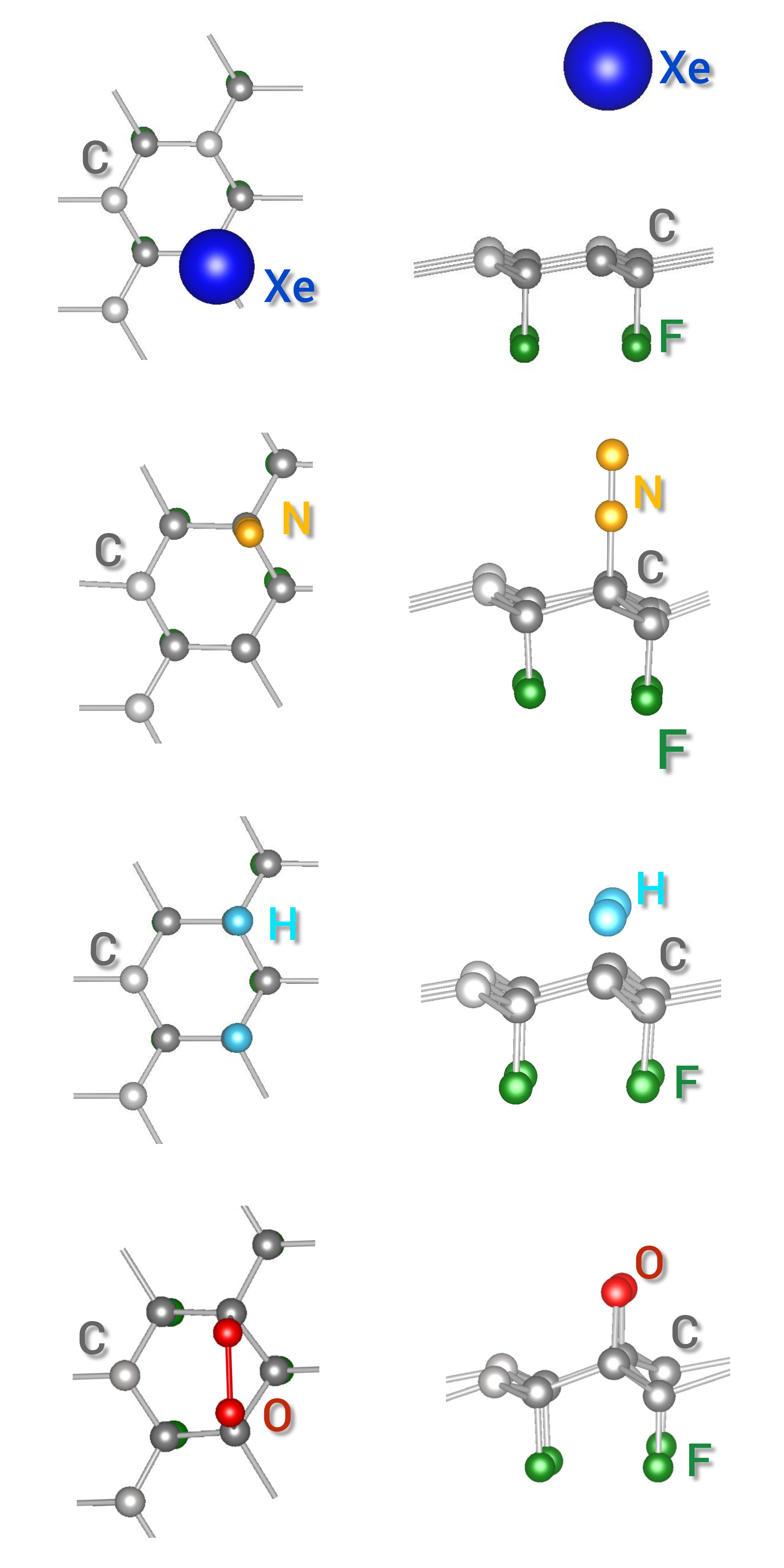}
\caption{\label{struct}
 Top and side views of the relaxed structures of the hFG sheet, 
 after adsorption of different chemical species.
}
\end{figure} 

\begin{table}
\caption{\label{bind}
 Obtained equilibrium parameters of the pristine (C$_8$) and 
 the half-fluorinated (\gf) graphene sheet after adsorption of various chemical species,
 including \f\ and \xe\ atoms and \nn, \oo, and \hh\ molecules;
 $E_b$ (eV): binding energy, 
 $2a$ (\AA): equilibrium lattice parameter of the supercell,
 $d$ (\AA): equilibrium molecule-sheet vertical distance.
 In two more important cases, a larger $4\times4$ supercell ($\super O_2$) is also considered.
 In these cases, the equilibrium lattice parameter is divided by two,
 to be comparable with the other results.
 It should be clarified that the binding energy and vertical distance
 have been calculated for the last chemical species written in 
 the first column.
}
\begin{ruledtabular}
\begin{tabular}{lccc}
   system        &  $E_b$  & $2a$  &  $d$  \\
\hline\vspace{-2mm}
\\
  C$_8$          &   ---   & 4.915 &  ---  \\
\\                        
  \gg/\f         & $-2.06$ & 4.927 & 1.492 \\
  \gg/\xe        & $-0.16$ & 4.914 & 3.920 \\
  \gg/\nn        & $-0.12$ & 4.923 & 2.953 \\
  \gg/\oo        & $-0.09$ & 4.914 & 2.650 \\
  \gg/\hh        & $-0.05$ & 4.917 & 2.808 \\
\\                         
  \gf/\f         & $-5.46$ & 5.079 & 1.420 \\
  \gf/\xe        & $-0.18$ & 5.070 & 3.764 \\
  \gf/\nn        & $-0.34$ & 5.085 & 1.395 \\
  \gf/\hh        & $-4.67$ & 5.095 & 1.106 \\
  \gf/\oo        & $-3.97$ & 5.071 & 1.334 \\
\\                        
  \gf/\f$_1$\nn  & $-0.29$ & 5.097 & 1.402 \\
  \gf/\ff\nn     & $+0.00$ & 5.118 & 1.409 \\
  \gf/\fff\nn    & $-0.12$ & 5.167 & 3.568 \\
\\
  \gf/\f$_1$\oo  & $-3.49$ & 5.124 & 1.324 \\
  \gf/\ff\oo     & $-3.27$ & 5.174 & 1.332 \\
  $\super O_2$   & $-2.38$ & 5.158 & 1.307 \\
\\
  \gf/\ff\oh     & $-4.44$ & 5.180 & 1.399 \\\vspace{1mm}
  \gf/\ff\of     & $-2.29$ & 5.212 & 1.333 \\
\\
  $\super(OF)_2$ & $-3.08$ & 5.179 & 0.961 \\
\end{tabular}
\end{ruledtabular}
\end{table}

\begin{figure}
 \includegraphics[scale=0.88]{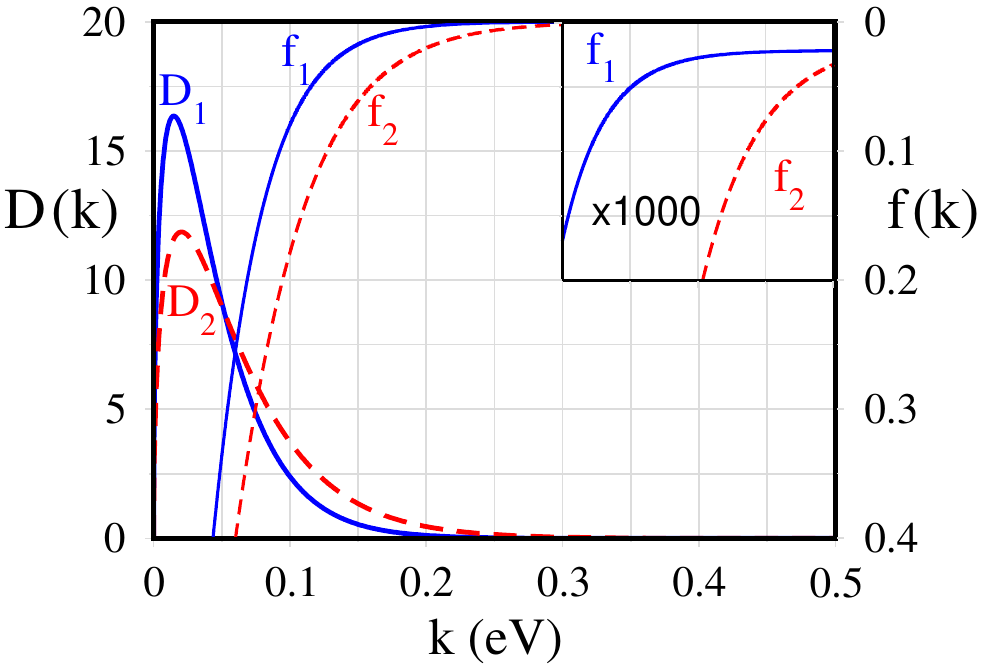}
\caption{\label{maxbolt}
 Maxwell-Boltzmann distribution D(k) as a function
 of the particle kinetic energy k at 70\deg\ (D$_1$) and 200\deg\ (D$_2$),
 and the resulting fraction of particles f(k) with 
 a kinetic energy above k ($f(k)=\int_k^{\infty}D(k')dk'$).
 The inset shows the enlarged tail of the f(k) functions.
}
\end{figure}

To verify this scenario, we considered the adsorption of four 
inert chemical species (Xe atom and \hh, \nn, and \oo\ molecules)
on the pristine and the half-fluorinated graphene sheet. 
These natural molecules are available in the dry atmosphere of a glove box used for 
the traditional fluorination of Gr samples.
The adsorption of these species was studied in a 2$\times$2 lateral 
supercell (Fig.~\ref{struct}) to reduce the effects of the adjacent replicas. 
The obtained relaxed geometries are sketched in Fig.~\ref{struct} and 
the computed physical parameters are presented in table~\ref{bind}.
We observe the physical adsorption of these inert species on 
the pristine sheet (C$_8$),
the corresponding absolute adsorption energies are below 0.2\,eV,
whereas the bond lengths are above 2.6\,\AA.
In the case of the heavy Xe atom, we considered the spin-orbit correction 
and found a negligible relativistic effect in the corresponding binding energy. 
These observed low binding energies may be well compensated  
by the room temperature vibrational energy of the adsorbate atoms
as well as by the kinetic energy of the incident F atoms during the fluorination process.
To speculate the vibrational energy of the adsorbates,
we calculated the gamma point vibrational modes of the xenon atom,
adsorbed on the pristine (C$_8$) sheet.
The results indicate three soft modes at about 6, 10, and 31\,cm$^{-1}$,
all below 4\,meV.
These soft modes are well saturated at room temperature and hence
the corresponding vibrational energy would be $3k_BT$,
which equals about 0.08\,eV at room temperature.
Similarly, a diatomic molecule physically adsorbed on graphene,
is expected to have five soft vibrational modes and hence the corresponding 
room temperature vibrational energy should be roughly about 0.13\,eV.
Adapting a Maxwell-Boltzmann distribution for the incident F atoms (Fig.~\ref{maxbolt}),
we observe that at the temperature of 70\deg\, about 8\% of the F atoms
have a kinetic energy higher than 0.1\,eV.
Therefore, we conclude that the thermal vibrational energies and 
the kinetic energy of the incident F atoms at 70\deg\ are sufficient to clean up
the surface of the pristine graphene from the weakly bounded chemical contaminants
(table~\ref{bind}) and consequently give rise to a fast half-fluorination process 
for the pristine graphene at 70\deg.

The obtained binding energies on hFG (\gf)
exhibit a clear enhancement with respect to the corresponding values on the pristine graphene.
This observation confirms the proposed scenario about the chemical activation 
of the second facet of graphene after half-fluorination.
The largest activation is seen for the \oo\ and \hh\ molecules where
experience a tight chemical bonding to the half-fluorinated graphene.
As it is seen in Fig.~\ref{struct}, both molecules prefer a parallel
configuration to the sheet. In the case of \hh\ molecule,
the hydrogen atoms are almost decoupled on the sheet, 
and each one binds to its underlying carbon atom.
On the other hand, oxygen atoms keep their bonding on the graphene sheet
and thus slightly deviate from the carbon top sites.
Although \nn\ exhibits much weaker binding to \gf,
compared with \oo\ and \hh, it is still significantly enhanced
with respect to the \nn\ binding on the pristine graphene (C$_8$).
Moreover, this molecule prefers a vertical alignment on the sheet (Fig.~\ref{struct}).
The Xe atom shows the lowest binding enhancement, after half-fluorination
of graphene. The stable position of this noble element on \gf\ is 
the carbon top site, while on C$_8$ it prefers the hollow sites of the sheet.
Comparing the binding energy of F and other investigated species on \gf\ (table~\ref{bind})
clarifies the anticipated competition between these species to occupy
the adsorption sites of the second facet of the hFG.
In our proposed scenario, this competition is the main reason behind 
the very slow rate of the full-fluorination process of graphene.
It should be stressed that the main competition likely happens between oxygen molecules
and F atoms, because of the close binding energy of these entities.
Although \hh\ has closer binding energy to the fluorine atom, 
the very low amount of hydrogen in the environment rules out its significant impact
on the fluorination process.
However, our data suggest the presence of rare hydrogen defects in the fluorographene
samples which may have nontrivial influences on the electronic structure of the system.
It may also be noted that \nn\ and \hh\ exhibit a second stationary position 
on \gf\ which involves much weaker bonding relative to the first one discussed above.
In the case of \nn, this weaker bonding happens about 3.5\AA\ above the sheet
with a binding energy of $-0.11$\,eV, 
while, \hh\ displays a second stationary position around 2.73\,\AA\ 
above the sheet with a similar binding energy of $-0.11$\,eV.

Now, we can explain some important observations about graphene fluorination
in the dry atmosphere of a glove box.\cite{nair2010} 
It should be noted that the graphene sheets were assembled
on a gold grid to make both sides of the sheet exposable to F atoms.
However, one side of the sheet is practically more accessible to
the available adsorbates.
Hence, in the first stage, one side of graphene is rapidly fluorinated,
because available chemical species are not able to compete
with the fluorine atoms for occupying the adsorption site of the sheet.
The half-fluorination process significantly increases the chemical activity 
of the opposite facet of the sheet and hence in addition to the fluorine atoms,
mainly the oxygen molecules and to a lower extent,
the nitrogen molecules also occupy the adsorption sites of the hFG sheet 
and thus remarkably slow down the full-fluorination process.
Our scenario predicts very fast full-fluorination of graphene in an oxygen-free ambient,
confirmed by a very recent experiment.
Li and others observed very a high rate of fluorographene synthesis
at a vacuum condition of 10$^{-6}$\,mbar.\cite{li2020}
Moreover, the observation of the high chemical activity hfG rules out 
the stability and feasibility of the individual half-fluorinated graphene samples.

In the next step, we investigate the durability of \nn\ and 
then \oo\ impurities in the fluorographene samples.
The gamma point vibrational frequencies of an \nn\ molecule adsorbed
on C$_8$ and \gf\ were estimated to be (12,18,57,94,96) and (10,20,36,49,50) cm$^{-1}$,
respectively, which in both cases give rise to a thermal energy of 
about 0.15 and 0.20\,eV at 70 and 200\deg, respectively.
These thermal energies are not sufficient to compensate for the binding energy 
of an individual \nn\ molecule ($-0.34$\,eV) on hFG (table~\ref{bind}).
In order to provide more information,
we investigated the accumulation of F atoms and other chemical species in 
the neighborhood of an adsorbed \nn\ molecule on the hFG.
In the first step, we found that \nn\ molecule does not permit adsorption
of a further \nn\ in the adjacent sites.
Then, we studied the adsorption of F atoms in the neighborhood
of an \nn\ molecule (table~\ref{bind}).
This procedure was seen to weaken the binding of \nn\ molecule to the hFG.
Adsorption of one F atom (\gf/\f$_1$\nn) increases the binding energy of \nn\ molecule 
to $-0.29$\,eV while adsorption of two F atoms (\gf/\ff\nn) almost decouple
the \nn\ molecule from the sheet.
Adsorption of the third F atom shifts the molecule to its second
stationary position which is about 3.56\,\AA\ far above the sheet
and thus may be easily removed by thermal effects.
It should also be noted that an \nn\ adsorbate on \gf\ has 
a slight influence on the binding of F atoms in the adjacent sites.
The binding energy of the first, the second, and the third F atom
in the vicinity of \nn\ is $-5.41$, $-5.20$, and $-5.45$\,eV which are pretty close to
the binding energy ($-5.46$\,eV) of an individual F atom on the clean facet 
of the hFG (table~\ref{bind}).
These results show that while individual \nn\ adsorbates on hFG are rather stable,
the prior accumulation of F atoms in their vicinity detaches these molecules from
the surface, even at low temperatures of about 70\deg. 
 
\begin{figure}
\includegraphics[scale=0.75]{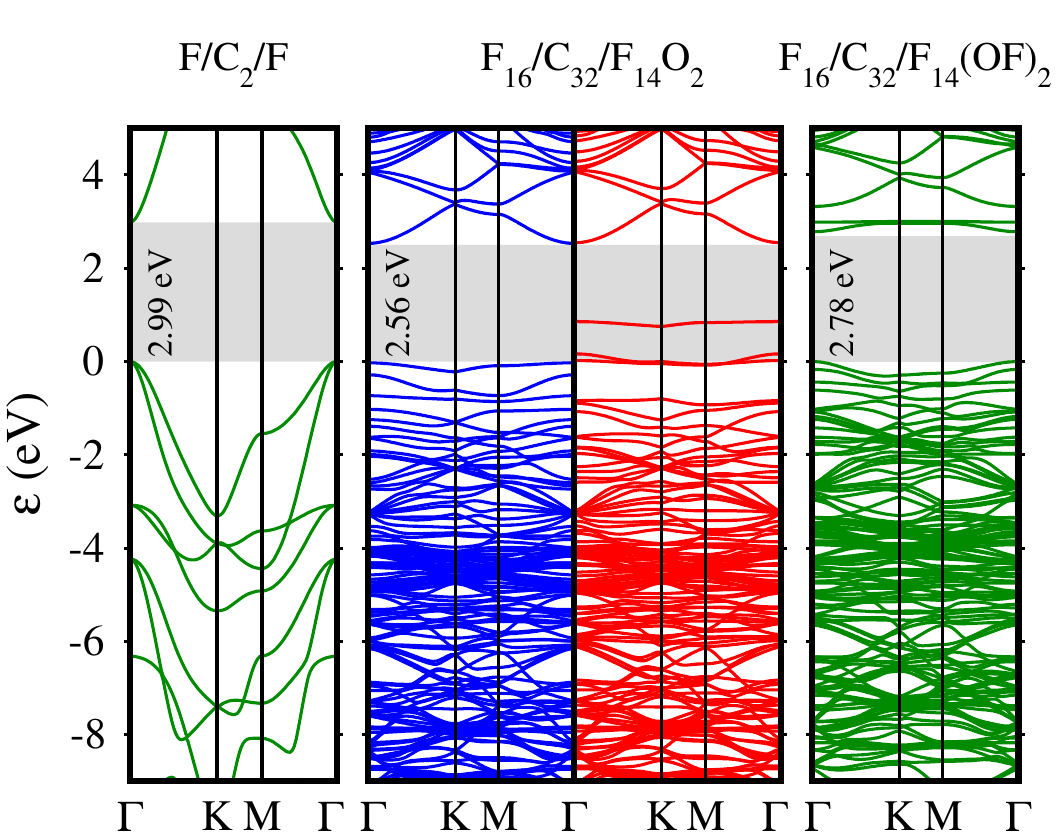}
\caption{\label{band}
 Obtained band structures for F/C$_2$/F (fluorographene), $\super O_2$,
 and $\super(OF)_2$. 
 $\super O_2$ exhibits an spin polarized band structure, 
 blue and red colors show spin majority and spin minority 
 states, respectively.
 The shaded areas show the band gap region and 
 the band gap values are written in the figures.
}
\end{figure}

The same scheme was applied around an \oo\ adsorbate on the hFG.
First, we found that this adsorbate significantly increases the binding energy 
of a further oxygen molecule in the adjacent sites to a value of about $-1.05$\,eV.
Then, adsorption of F atoms was considered around the \oo\ molecule.
Because of the parallel configuration of \oo\ on the hFG, 
only two F atoms may be adsorbed in the vicinity of \oo\ in a $2\times2$ supercell.
The first and the second F atoms increase the binding energy of oxygen 
to $-3.49$ and $-3.27$\,eV, respectively (table~\ref{bind}).
Despite these weakening effects, the \oo\ molecule is still tightly attached
to the sheet and thermal effects are far lower to compensate 
for the remained binding energy.
In order to reduce the effects of adjacent \oo\ molecules,
we repeated the last calculation in a larger $4\times4$ supercell
and found a high binding energy of about $-2.38$\,eV
for the \oo\ molecule fully surrounded by F atoms ($\super O_2$).
These findings argue that the complete elimination of the oxygen defects
from the fluorographene samples is an elaborating task.
This observation may explain the observed rare difference between 
the experimental (3.1-3.8\,eV)\cite{jeon2011} and the theoretical excitonic band gap 
of this system (5.1\,eV).\cite{yuan2015}
The obtained spin polarized band structure of $\super O_2$ is compared with 
that of an ideal FG sample in Fig.~\ref{band}.
It is seen that O$_2$ impurities may significantly decrease 
the theoretical band gap of FG, obtained within 
the single-particle Kohn-Sham framework.\cite{note1}
Moreover, some minority spin states occupy the band gap
and cross the Fermi level of the oxygen defected FG,
indicating the chemical activity of the oxygen defect in this system.

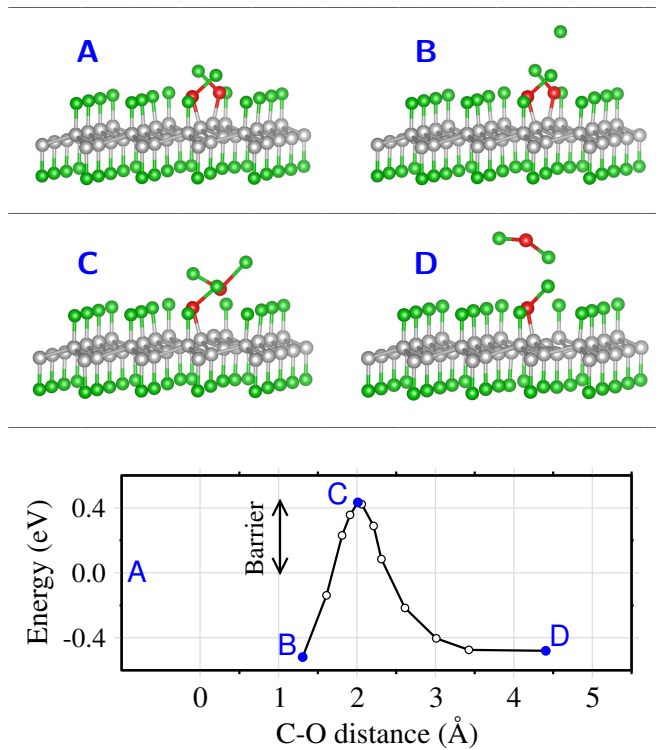
\begin{figure}
--------------------------------------------------------------------------------
 \includegraphics[scale=0.15]{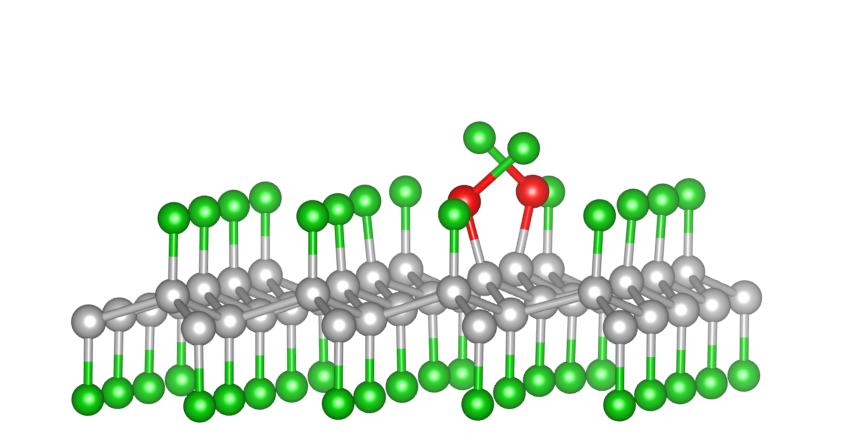}\includegraphics[scale=0.15]{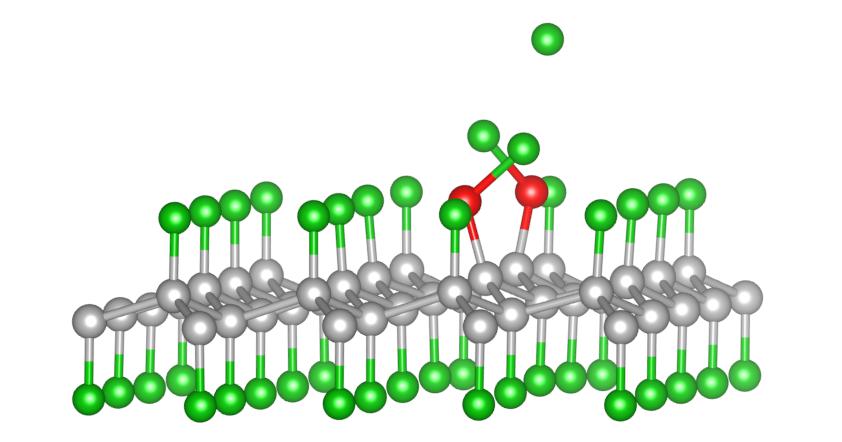}
--------------------------------------------------------------------------------
 \includegraphics[scale=0.15]{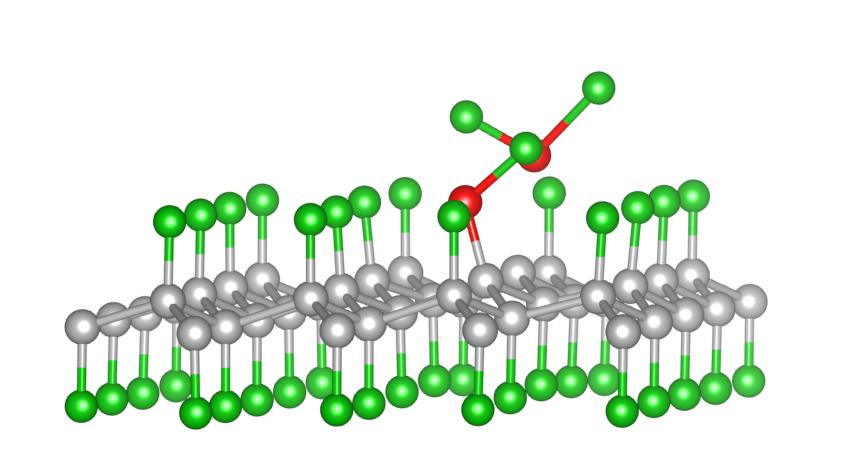}~~\includegraphics[scale=0.13]{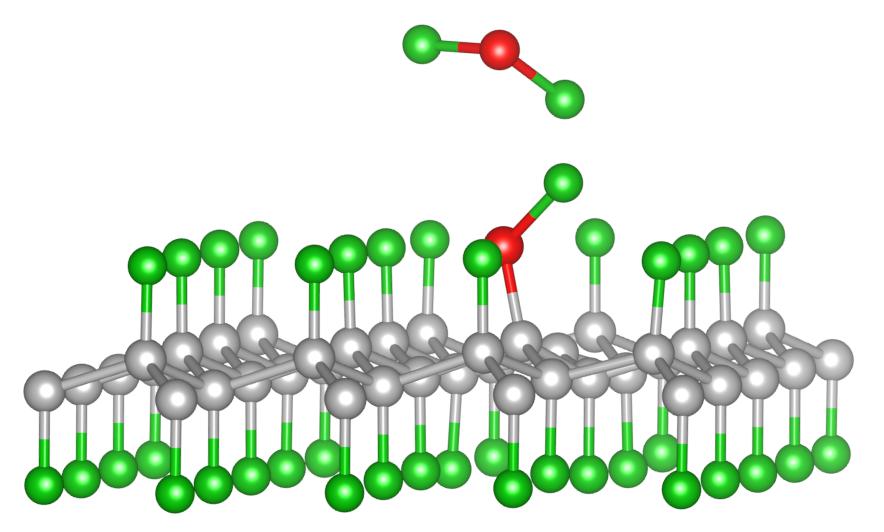}
--------------------------------------------------------------------------------
\begin{picture}(50,0)(0,0)
\put(-68,155){\makebox(,)[]{\textcolor{blue}{\large\bf\textsf A}}}
\put( 60,155){\makebox(,)[]{\textcolor{blue}{\large\bf\textsf B}}}
\put(-68,75){\makebox(,)[]{\textcolor{blue}{\large\bf\textsf C}}}
\put( 60,75){\makebox(,)[]{\textcolor{blue}{\large\bf\textsf D}}}
\end{picture}
 \includegraphics[scale=0.95]{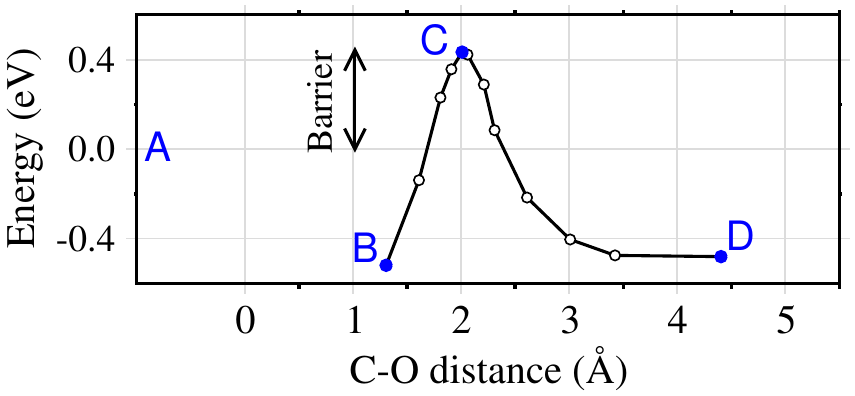}
\caption{\label{desorption}
 A: predicted atomic configuration of an oxygen defected FG 
    in a $4\times4$ supercell ($\super(OF)_2$).
    The red, green, and gray balls stand for the oxygen, fluorine, 
    and carbon atoms, respectively.
 B, C, D: initial, transition state, and final configurations of 
        the $\super(OF)_2+F$ complex in the proposed oxygen desorption reaction.
 bottom: optimized energy path of the oxygen desorption reaction
         as a function of the reaction coordinate.
}
\end{figure}

After confirming the very probable presence of O$_2$ defects in 
the FG samples, we considered their chemical activity 
by adsorbing a further \hh\ and \ff\ molecule on top of these defects. 
The obtained results, presented in table~\ref{bind}, 
exhibit the rather high chemical activity of this point defect.
The binding energy of \hh\ on \oo\ is about $-4.44$\,eV and 
we observed that this strong binding almost decouples 
the two H atoms from each other.
The binding of \ff\ on \oo\ is also predicted to be rather strong, 
each F atom attaches to one of the O atoms with a binding energy of 
about $-1.13$\,eV/atom.
Because of the much higher abundance of F atoms in the FG synthesis environment,
compared with the \hh\ molecules,
the anticipated \oo\ defects in FG are likely covered by further F atoms,
leading to the pairs of OF ligands.
In order to approach the more realistic situations,
we optimized the atomic configuration of these ligands in 
a larger $4\times 4$ supercell (Fig.~\ref{desorption})
and observed a cross geometry in the neighboring OF ligands.
The electronic band structure of this system is also calculated and 
presented in Fig.~\ref{band}.
Compared with the system with a bare oxygen defect ($\super O_2$), we observe that
the semiconducting feature of FG is well recovered with a band gap
of about 2.78\,eV, that it is still about 7\% lower than the ideal FG.

In the last part of the paper, we provide some first-principles arguments 
about the significant enhancement of the full-fluorination procedure 
of graphene in the glove box, after increasing the temperature 
from 70\deg\ to 120\deg\ and 200\deg.\cite{nair2010,creutzburg2021}
According to the above discussions, neighboring OF ligands
are the most likely defects in the FG samples,
hence the full-fluorination process requires an effective mechanism
for removing the adsorbed oxygen atoms.
We focus on the optimized model of an oxygen defected FG sample in 
a $4\times 4$ supercell (Fig.~\ref{desorption}) and propose 
the following oxygen desorption reaction:
$$ \super(OF)_2 + F ~\longrightarrow~ \super(OF) + F_2O $$
An incident F atom picks up one of the OF ligands from the sample and 
forms a free F$_2$O molecule.
In order to calculate the energy barrier of this reaction,
we selected the vertical distance between the oxygen atom of 
the target ligand and its underlying carbon atom (C-O distance) as 
the reaction coordinate and then optimized the energy of the system
at different values of this coordinate.
The obtained reaction energy path is presented in Fig.~\ref{desorption}.
Four configurations are labeled in the energy path and sketched in the figure
for a better understanding of the oxygen desorption reaction.
As it was mentioned, the configuration A represents the predicted structure of 
the oxygen defected FG in a $4\times 4$ supercell.
The minimized energy of this system plus the energy of a free F atom
has been selected as the reference of energy in the reaction energy path.
The configuration B shows an equilibrium position of the incident F atom above the ligand,
however, the kinetic energy of the atom prevents it from stopping at this position.
Hence, the incident F atom may approach and then attach to the target OF ligand
to form an F$_2$O molecule on the surface.
If the incident F atom has enough kinetic energy, the formed molecule
at the surface may climb the energy barrier to reach the transition configuration C,
where the oxygen atom of the V-shape F$_2$O molecule is about 2\,\AA\ 
above the central graphene layer (Fig.~\ref{desorption}).
On the other hand, in the final configuration D, the desorbed F$_2$O molecule
exhibits an inverted V-shape geometry.
Our results indicate an energy barrier of about 0.43\,eV for 
the proposed oxygen desorption reaction.
It should be noted that the activation barrier should be calculated from A,
not B that is an intermediate configuration of the reaction.

In our proposed oxygen desorption reaction, the incident F atom should have 
a kinetic energy of at least 0.43\,eV, to overcome the activation barrier of the reaction.
According to the Maxwell-Boltzmann distribution (Fig.~\ref{maxbolt}),
at 70\deg\, only about $2.5\times 10^{-3}$\,\% of the particles have 
a kinetic energy above 0.43\,eV, which may explain
the very low rate of the full-fluorination of graphene 
at this temperature.
Nair and others found that complete fluorination of a graphene sheet at 70\deg\
takes more than two weeks, while at 200\deg\ about 10\,hours is enough for 
the full-fluorination procedure.\cite{nair2010}
At 200\deg, the number of particles with enough kinetic energy increases
by a factor of about 4 (Fig.~\ref{maxbolt}), 
which qualitatively confirms the mentioned rate enhancement.
However, a more accurate comparison of the reaction rate ($k(T)$) 
at different temperatures requires the Arrhenius scheme:
$$k(T)=Ae^{-E_a/k_BT}$$
$A$, $E_a$, $k_B$, and $T$ are frequency factor, activation energy,
the Boltzmann constant, and the Kelvin temperature, respectively.
Inserting the obtained activation energy (0.43\,eV) and 
neglecting the temperature dependence of the frequency factor,
the above equation predicts that the reaction rate at 200\deg\
is about 55 times higher, compared with 70\deg.
Hence the more than two weeks (336\,hours) full-fluorination procedure
of graphene at 70\deg, is expected to reduce to more than 6\,hours at 200\deg,
in close agreement with the observation of Nair and colleagues 
($\sim$10\,hours).\cite{nair2010}
In the same way, the full-fluorination time at 120\deg\ is predicted to be 
more than 39\,hours, well consistent with the very recent record 
of Creutzburg et al. (48\,hours).\cite{creutzburg2021}
Therefore, our proposed oxygen defected FG and oxygen desorption mechanism
explain well the observed features in the graphene fluorination procedure. 

\section{Conclusions}

In this work, we performed extensive full-potential DFT calculations on 
various arrangements of the relevant atomic and molecular defects in fluorographene (FG)
to understand the experimentally observed two stages fluorination procedure of Gr.
In a glove box, at 70\deg\, the half-fluorination of a pristine graphene sheet 
was found to happen quickly while 
the full-fluorination procedure takes more than two weeks.
In the first step, the adsorption energy profile of a fluorine atom on the pristine and
half-fluorinated graphene (hFG) sheet was calculated and compared to identify
a significant enhancement in the chemical activity of graphene after 
the half-fluorination stage.
This observation rules out the feasibility of individual half-fluorinated graphene samples. 
Afterward, adsorption of the relevant chemical species, including Xe atom
and \nn, \oo, and \hh\ molecules were considered on the pristine
and half-fluorinated graphene to confirm the predicted chemical activation.
After considering various arrangements of the adsorbed \nn\ molecule on FG,
it was argued that this point defect is not durable and typical 
thermal energies at 70\deg\ are enough to clean up the FG samples from this defect.
On the other hand, the \oo\ molecule exhibits a different behavior and may
form thermally stable defects in FG.
We calculated diverse configurations of this molecule and concluded
that pairs of OF ligands with a cross geometry are likely the most abundant impurities in 
the realistic FG samples, opposing the full-fluorination procedure.
We proposed an oxygen desorption reaction for removing the OF ligands from the sample
and thus enhancing the complete fluorination of graphene.
In this mechanism, an incident F atom with enough kinetic energy attaches
to an OF ligand to form an F$_2$O molecule on the sheet.
This molecule then should overcome an energy barrier of about 0.43\,eV to leave the sample.
These results explain very well the substantially increased rate of 
the full-fluorination procedure at 120 and 200\deg.

\section{ACKNOWLEDGMENTS}
The authors appreciate the useful comments of Dr. Mehdi Abdi 
from Isfahan University of Technology (IUT).
This work was supported by the IUT Vice-Chancellor in Research Affairs.

\bibliography{fluorographene}

\begin{thebibliography}{17}
\bibitem[{not()}]{note1}
\bibinfo{note}{Unfortunately, we were not able to verify this trend in the
  excitonic or quasiparticle band gaps, beacuse our computational resources are
  not powerfull enough to apply first principles GW or excitonic calculations
  to our rather large supercells.}
\end{thebibliography}

\end{document}